\documentclass{llncs}
\usepackage{makeidx}  % allows for indexgeneration
\begin{document}
\frontmatter          % for the preliminaries
\pagestyle{headings}  % switches on printing of running heads
%\addtocmark{Hamiltonian Mechanics} % additional mark in the TOC
%
%\chapter*{Preface}
%
%\ldots

%\vspace{1cm}
%\begin{flushright}\noindent
%\ldots \hfill \ldots\\
%\ldots \\
%\ldots
%\end{flushright}
%
%\chapter*{Organization}
%\ldots
%
%\section*{Executive Commitee}
%\begin{tabular}{@{}p{5cm}@{}p{7.2cm}@{}}
%\ldots
%\end{tabular}
%
%\section*{Program Commitee}
%\begin{tabular}{@{}p{5cm}@{}p{7.2cm}@{}}
%Conference Chair:&Ole Lehrmann Madsen (\AA rhus University, DK)\\
%Program Chair:   &Walter Olthoff (DFKI GmbH, Germany)\\
%Organizing Chair:&J\o rgen Lindskov Knudsen (\AA rhus University, DK)\\
%Tutorials:&Birger M\o ller-Pedersen\hfil\break
%(Norwegian Computing Center, Norway)\\
%Workshops:&Eric Jul (University of Kopenhagen, Denmark)\\
%Panels:&Boris Magnusson (Lund University, Sweden)\\
%Exhibition:&Elmer Sandvad (\AA rhus University, DK)\\
%Demonstrations:&Kurt N\o rdmark (\AA rhus University, DK)
%\end{tabular}
%
%\begin{multicols}{3}[\section*{Referees}]
%\ldots
%\end{multicols}
%
%\section*{Sponsoring Institutions}
%
 %\ldots
%
%\tableofcontents
%
\mainmatter              % start of the contributions
\title{Mandelbrot's $1/f$ fractional renewal models of 1963-67: The non-ergodic missing link between change points and long range dependence.}
\titlerunning{Hamiltonian Mechanics}  % abbreviated title (for running head)
%                                     also used for the TOC unless
%                                     \toctitle is used
%
\author{Nicholas Wynn Watkins\inst{1}\inst{2}\inst{3}\inst{4}\inst{5}}
\authorrunning{Ivar Ekeland et al.} % abbreviated author list (for running head)
%
%%%% list of authors for the TOC (use if author list has to be modified)
%\tocauthor{Ivar Ekeland, Roger Temam, Jeffrey Dean, David Grove,
%Craig Chambers, Kim B. Bruce, and Elisa Bertino}
%
\institute{Centre for the Analysis of Time Series, London School of Economics, London, UK,\\
\email{N.Watkins2@lse.ac.uk},
%\\ WWW home page:\texttt{...}
\and Faculty of Mathematics, Computing and Technology, Open University, Milton Keynes, UK,\\
\and
Centre for Fusion, Space and Astrophysics, University of Warwick,UK\\
\and
Universit{\"a}t Potsdam, Institut f{\"u}r Physik und Astronomie, Campus Golm, Potsdam-Golm, Germany
\and
Max Planck Institute for the Physics of Complex Systems, Dresden, Germany.
}

\maketitle              % typeset the title of the contribution

\begin{abstract}
The problem of  1/f noise has been with us for about a century. Because it is so often framed in Fourier spectral language, the most famous solutions have tended to be the stationary long range dependent (LRD) models such as Mandelbrot's fractional Gaussian noise.   In view of the increasing importance to physics of non-ergodic fractional renewal models, I present preliminary results of my research into the history of Mandelbrot's very little known work in that area from 1963-67. I speculate about how the lack of awareness of this work in the physics and statistics communities may have affected the development of complexity science; and I discuss the differences between the Hurst effect, 1/f noise and LRD, concepts which are often treated as equivalent.
\end{abstract}
\section{Ergodic and non-ergodic solutions to the paradox of 1/f noise}
``The problem of 1/f noise" has been with us for about 100 years since the pioneering work of Schottky and Johnson \cite{Grigolini2009,Graves2014b,Watkins2016}. It is usually framed as a spectral paradox, i.e.``how can the Fourier spectral density $S'(f)$  of a stationary process take the form $S'(f) \sim 1/f$ and thus be singular at the origin (or equivalently how can the autocorrelation function ``blow up" at large lags and thus not be summable) ? ". When a problem is seen this way, the solution will also tend to be sought   in spectral terms. The desire of  for  a solution to the problem with a satisfying level of generality  increased in the 1950s with the recognition of an analogous time domain effect (the Hurst phenomenon) seen in the statistical growth of range in Nile minima \cite{Graves2014b}.  The first stationary solution which could exhibit the Hurst effect and $1/f$ noise was presented by Mandelbrot in 1965 using fractional Gaussian noise (fGn), the increments of fractional Brownian motion (fBm), and was subsequently developed with Van Ness and Wallis, particularly in the hydrological context \cite{Graves2014b,Mandelbrot2002}. fGn is  a stationary ergodic process, for which a power spectrum is a natural and well-defined concept, the paradox here resides in its singular behaviour at zero.
  
However, in the last two decades it has increasingly been realised in physics that another class of models, the fractional renewal processes, can also give $1/f$ noise in a very different way\cite{Margolin2006}. Physical interest has come from phenomena such as weak ergodicity breaking (in e.g. blinking quantum dots \cite{Stefani2009,Niemann2013,Sadegh2015}) and the related question of how many different classes of model can share the common property of the 1/f spectral shape (e.g. \cite{Rodriguez2014,Rodriguez2015}). In view of this resurgence of activity, my first aim in this paper  is to report (Section 2) preliminary results from my historical research which has found, to my great surprise, that the dichotomy between ergodic and non-ergodic origins for $1/f$ spectra was not only recognised but also published about by Mandelbrot 50 years ago in some  still remarkably little known work  \cite{BergerMandelbrot1963,Mandelbrot1965a,Mandelbrot1965b,Mandelbrot1967}. He carried it out in parallel with his seminal work on the ergodic, stationary fGn model.  In these papers, and the bridging essays he wrote when he revisited them late in life for his collected Selecta volumes, particularly \cite{Mandelbrot1999,Mandelbrot2002}, he developed and published a series of  fractional renewal models. In these the periodogram, the empirical autocorrelation function (acf), and the observed waiting time distributions, all grow in extent with the length of time over which they are measured. He explicitly \cite{Mandelbrot1967} drew attention to this non-ergodicity and its origins in what he called ``conditional" stationarity. He explicitly contrasted the fractional renewal models with the stationary, ergodic fGn which is today very much better known to physicists, geoscientists and many other time series analysts \cite{Beran2013,Graves2014b}. Mandelbrot's work at IBM was itself in parallel with other developments,  one notable example being the work of Pierre Mertz \cite{Mertz1961,Mertz1965} at RAND on modelling telephone errors,  so my preliminary report is not an attempt to assign priority. I hope to return to the history of this period in more detail in future articles. 

My  next purpose (Section 3) is  to clarify the subtle differences between 3 phenomena: the empirical Hurst effect, the appearance of $1/f$ noise in periodograms, and the concept of LRD as embodied in the stationary ergodic fGn model, and to set out their hierarchy  with respect to each other, aided in part by this  historical perspective. This paper will not deal with another possibility-multiplicative models-\cite{Mandelbrot1999,Rodriguez2014}, though I do of course recognise that they remain a very important alternative source of $1/f$ behaviour, particularly that arising from turbulent cascades. I will also not be considering $1/f$-type spectra arising from nonstationary self-similar walks such as fractional Brownian motion.

I will (Section 4) conclude by speculating on the how the relative neglect of\cite{BergerMandelbrot1963,Mandelbrot1965a,Mandelbrot1965b,Mandelbrot1967} at the time of their publication may have had long-term effects.

\subsection{fGn and the fractional renewal process compared}
fGn \cite{Beran2013} is effectively a derivative of fractional Brownian motion $Y_{H,2}(t)$:
\begin{equation}
Y_{H,2}(t) = \frac{1}{C_{H,2}} \int_R dL_2 (s) \ K_{H,2}(t-s) 
\end{equation} 
which in turn extends the Wiener process to include a self-similar, memory kernel  $K_{H,2} (t-s)$,   where 
\begin{equation}
K_{H,2}(t-s) =  \large[ (t-s)_{+}^{H-1/2} - (-s)^{H-1/2} \large]  
\end{equation}
thus giving a decaying, non-zero weight to all of the values in the time integral over $dL$.

 In consequence fGn shows long range dependence, and has indeed become a very important paradigmatic model for LRD. The attention paid to its 1/f spectrum, and long-tailed acf, as diagnostics of LRD, has often led to it being forgotten that its stationarity  is an equally essential ingredient for LRD in this sense. Intuitively one can see that without stationarity there can be no LRD because there is no infinitely long past history over which the process can be dependent. Models like fGn, and also fractionally integrated noise (FIN) and the autoregressive fractionally integrated moving average (ARFIMA) process,  which have been widely studied in the statistics community (e.g. \cite{Beran1994,Beran2013}) exhibit LRD  {\em by construction}, i.e. stationarity is assumed at the outset in defining them.  

While undeniably important to time series analysis and the development of complexity science, we can already see from the restriction to stationary processes that the  LRD concept, at least as embodied in fGn, will be insufficient to describe the whole range of either $1/f$ or Hurst behaviour that observations may present us with. Full awareness of this limitation has been slow because of three  widespread, deeply-ingrained, but unfortunately erroneous beliefs: i) that an observed Fourier periodogram can {\em always} be taken to estimate a power spectrum, ii) that the Fourier transform of an empirically obtained periodogram is {\em always} a meaningful estimator of an autocorrelation function, and iii) that the  observation of a 1/f Fourier periodogram  in a time series {\em must} imply the kind of long range dependence that is embodied in the ergodic fractional Gaussian noise model.  The first two beliefs are routinely cautioned against in any good course or book on time series analysis, including classics like Bendat's \cite{Bendat1958}. The third belief remains highly topical, however, because it is only relatively recently being appreciated in the theoretical physics literature just how distinct two paradigmatic classes of 1/f noise model are, and how these differences relate not only to LRD but also to the fundamental physical question of weak ergodicity breaking (e.g. \cite{Bouchaud1992,Margolin2006,Niemann2013}).

The second paradigm for 1/f noise mentioned above is the fractional renewal class, which  is a descendent of the classic random telegraph model \cite{Bendat1958}, and so looks at first sight to be stationary and Markovian, but has switching  times at power law distributed intervals. A particularly well studied variant is the alternating fractal renewal process (AFRP, e.g. \cite{LowenTeich1993,LowenTeichBook}), which is also closely connected to  the renewal reward process in mathematics. When studied in the telecommunications context, however, the AFRP has often had a cutoff applied to its switching time distribution for large times to allow analytical tractability. The use of an upper cutoff unfortunately masks some of its most physically interesting behaviour, because when the cutoffs are not used  the periodogram, the empirical acf, and {\em observed} waiting time distributions, all {\em grow} with the length of time over which they are measured, rendering the process both non-ergodic  and non-stationary in an important sense (Mandelbrot preferred his own term ``conditionally stationary").  In particular, Mandelbrot stressed that the   process no longer obeys the necessary conditions on the Wiener-Khinchine theorem for its empirical periodogram to be interpreted as an estimate of the power spectrum. This property of weak ergodicity breaking (named by Bouchaud in the early 1990s \cite{Bouchaud1992}) is now attracting much interest in physics,  see e.g.  Niemann et al \cite{Niemann2013},  on the resolution of the low frequency cutoff paradox, and  subsequent developments \cite{Leibowich2015,Dechant2015,Rodriguez2014,Rodriguez2015}.
 
The existence of this alternative, nonstationary, nonergodic fractional renewal  model makes it clear that there is a difference between the observation of an empirical 1/f noise alone, and the presence of the type of LRD that is embodied in the stationary ergodic fGn model. We will develop this point further in section 3, but  will first go back to the  1960s and Mandelbrot's twin tracks to $1/f$.
%%%%%%%%%%%%%%%%%%%%%%%%%%%%%%%%%%%%%%%%%%%%%%%%%%%%%%%%%%%%%%%%%%%%%%

\section{Mandelbrot's fractional renewal route to ``1/f"} 
What seems to have gone almost completely unnoticed,  is the remarkable fact that Mandelbrot was not only aware of the distinction between fGn and fractional renewal models \cite{Mandelbrot1999,Mandelbrot2002}, but also published a nonstationary model of the AFRP type in 1965 \cite{Mandelbrot1965a,Mandelbrot1965b} and had explicitly discussed the time dependence of its power spectrum as a symptom on non-ergodicity by 1967 \cite{Mandelbrot1967}.  

There are 4 key papers in Mandelbrot's consideration of fractional renewal models. The first, cowritten with physicist Jay Berger  \cite{BergerMandelbrot1963}, appeared in  IBM Journal of Research and Development. It dealt with errors in telephone circuits, and its key point point was the power law distribution of times between errors, which were themselves assumed to take discrete values. Switching models were already being looked at, and the authors acknowledged that  Pierre Mertz of RAND had already studied a power law switching model \cite{Mertz1961}, but Mandelbrot's early exposure to the extended central limit theorem, and the fact that he was studying heavy tailed models in economics and neuroscience among other applications, evidently  helped him to see their broader significance.  

The second, \cite{Mandelbrot1965a} was in the IEEE Transactions on Communication Technology, and essentially also used the model of Berger and Mandelbrot. The abstract makes it clear that it describes:   
\begin{quote}
{\it ... a model of certain
random perturbations that appear to come in clusters, or bursts. This will be achieved by introducing the concept of ``self-similar
stochastic point process in continuous time." The resulting mechanism presents fascinating peculiarities from the mathematical
viewpoint. In order to make them more palatable as well as to help in the search for further developments, the basic concept of ``conditional
stationarity" will be discussed in greater detail than would be strictly necessary from the viewpoint of the immediate engineering
problem of errors of transmission. }
\end{quote}

It is clear that by 1965 Mandelbrot had come to appreciate that the application of the Fourier periodogram to the fractional renewal process would give ambiguous results, saying in \cite{Mandelbrot1965a} that: 
\begin{quote}
{\it The now classical technique of spectral analysis is inapplicable to the processes examined in this paper but it is
sometimes unavoidable that otherwise excellent spectral estimates be applied in this context. Another publication of
the author[that paper's Ref 18] is devoted to an examination of the expected outcomes of such operations. This will lead to fresh concepts
that appear most promising indeed in the context of a statistical study of turbulence, excess noise, and other
phenomena when interesting events are intermittent and bunched together (see also [that paper's Ref 19]).}
\end{quote}

The  third key paper, the ``other publication ... Ref 18", resulted from an IEEE conference talk in 1965. It  \cite{Mandelbrot1965b} is now available but in the {\em post hoc} edited form in which all his papers appeared in his Selecta\cite{Mandelbrot1999,Mandelbrot2002}. ``Reference 19", meanwhile, seems originally to have been intended to be a paper in the physics literature, the fate of which is not clear to me but whose role was effectively taken over by the fourth key paper \cite{Mandelbrot1967}. With the proviso that the Selecta version of  \cite{Mandelbrot1965b}  may not fully reflect the original's content, one can nonetheless see that by mid-1965 Mandelbrot was already focusing on the implications for ergodicity of the conditional stationarity idea.  He remarked that:
\begin{quote} 
{\it In other words, the existence of $f^{\theta-2}$ noises challenges the mathematician to reinterpret spectral measurements otherwise than in ``Wiener-Khinchin" terms.
[...] operations meant to measure the Wiener-Khinchin spectrum may unvoluntarily measure something else, to be referred to as the ``conditional spectrum" of a ``conditionally covariance stationary" random function.}
\cite{Mandelbrot1967}
\end{quote}
Taking the two papers \cite{Mandelbrot1965b,Mandelbrot1967}  together we can see that  Mandelbrot expanded on this vision by discussing several fractional renewal models,  including in \cite{Mandelbrot1965b}  a three state, explicitly nonstationary model with waiting times whose probability density function decayed as a power law $p(t) \sim t^{-(1+\theta)}$ .  This   stochastic process was intended as a ``cartoon" to model intermittency, in which ``off" periods of no activity were interrupted by jumps to a negative (or positive) ``on" active state. His key finding, confirmed in \cite{Mandelbrot1967} for a model with an arbitrary number of discrete levels,  was that the traditional Wiener-Khinchine spectral diagnostics would return a   $1/f$ periodogram and thus a spectral ``infrared catastrophe" when viewed with traditional methods, but building on the notion of conditional stationarity proposed in \cite{Mandelbrot1965a}, that a conditional power spectrum $S(f,T)$ could be decomposed into a  stationary part in which no catastrophe was seen, and one  depending on the time series' length $T$, multiplying a slowly varying function $L(f)$.  
He found
\begin{equation}
S(f,T) \sim f^{\theta-1}L(f)Q(T)
\end{equation}
where $Q(T)T^{1-\theta}$ was slowly varying, and that the conditional spectral density $S'(f,T)$ obeys
\begin{equation}
S'(f,T) = \frac{d}{df}S(f,T)  \sim f^{\theta-2} T^{\theta-1}L(f)
\end{equation}
Rather than representing a true singularity in power at the lowest frequencies,  in the Selecta \cite{Mandelbrot1999} he described the apparent  infrared catastrophe in the power spectral density in the fractional renewal models as a ``mirage" resulting from the fact that the moments of the  model varied in time in a step-like fashion, a property he called ``conditional covariance stationarity".

In \cite{Mandelbrot1967} Mandelbrot noted a clear contrast between his conditionally stationary, non-Gaussian fractional renewal
$1/f$ model and his stationary Gaussian fGn model (the 1968 paper concerning which, with Van Ness, was then in press at SIAM Review):
\begin{quote} 
{\it Section VI [... of this paper... ] showed that {\em some} $f^{\theta-2}L(f)$ noises have a very erratic sampling behavior. Some {\em other} $f^{\theta-2}$ noises are Gaussian and, therefore, perfectly ``well-behaved;" an example is provided by the ``fractional white noise" [i.e. fGn] which is the formal derivative of the process of Mandelbrot and Van Ness [i.e. fBm]}
%\cite{Mandelbrot1967}
\end{quote}

He identified the origin of this erratic sampling behaviour in the non-ergodicity of the fractional renewal processes. Niemann et al \cite{Niemann2013} have recently given a very precise analysis of the behaviour of the random prefactor $S(T)$ , obtaining its Mittag-Leffler distribution and checking it by simulations.

\section{The Hurst effect vs. 1/f vs. LRD}

Informed in part by the above historical investigations, the purpose of this section is now to distinguish conceptually between 3 phenomena which are still frequently elided.

To recap, the phenomena are:
\begin{itemize}
\item The Hurst effect: the {\em observation} of ``anomalous" growth of range in a time series using a diagnostic such as Hurst and Mandelbrot's $R/S$ or detrended fluctuation analysis (DFA)(e.g. \cite{Graves2014b,Beran2013}).
\item $1/f$ noise: the {\em observation} of  singular low frequency behaviour in the empirical periodogram of a time series.  
\item Long range dependence (LRD): a property of a stationary model {\em by construction}. This  can only be {\em inferred} to be a property of an empirical time series if certain additional conditions are known to be met, including the important one of stationarity 
\end{itemize}

The reason why it is necessary to unpick the relationship between these ideas is that there are  three commonly held misperceptions about them.  

\paragraph{The first is that observation of the Hurst effect in a time series necessarily implies stationary LRD.} This is ``well known" to be erroneous, see e.g. the work of  \cite{Bhattacharya1983}  who showed the Hurst effect arising from an imposed trend rather than from stationary  LRD, but is nonetheless in practice still not very widely appreciated. 

\paragraph{The second is that observation of the Hurst effect in a time series necessarily implies a $1/f$ periodogram.} Although less ``well known", \cite{FranzkePNAS}, for example, have shown an example where the Hurst effect arose in the Lorenz model which has an exponential power spectrum rather than $1/f$.  

\paragraph{The third is the idea that observation of a $1/f$ periodogram necessarily implies stationary LRD.} As noted above, this is  a more subtle issue, and although little appreciated since the pioneering work of \cite{Mandelbrot1965a,Mandelbrot1965b,Mandelbrot1967} it has now become central to the investigation of weak ergodicity breaking in physics.

\subsection{The Hurst effect}
The Hurst effect was originally observed as the growth of range in a time series, at first the Nile. The original diagnostic for this effect was R/S. Using the notation $J$ (not $H$) for the Joseph (i.e. Hurst) exponent that Mandelbrot latterly advocated \cite{Mandelbrot2002}, the Hurst effect is seen when the rescaled range\cite{Beran2013,Graves2014b} grows with time as
\begin{equation}
\frac{R}{S}\sim \tau^J
\end{equation}
in the case that $J \ne 1/2$. During the  period between Feller's proof that an iid stationary process had $J=1/2$, and Mandelbrot's  papers of 1965-68 on long range dependence in fGn \cite{Graves2014b}, there was a controversy about whether the Hurst effect was a consequence of nonstationarity and/or a pre-asymptotic effect.  This controversy has never fully subsided \cite{Graves2014b} because Occam's Razor frequently favours at least the possibility of change points in an empirically determined time series (e.g. \cite{Mikosch}), and because of the (at first sight surprising) non-Markovian property of fGn. 

A key point to appreciate is that it is easier to generate the Hurst effect over a finite scaling range, as measured for example by $R/S$,  than a true wideband 1/f spectrum. \cite{FranzkePNAS} for example shows how a Hurst effect can appear over a finite range even when the power spectrum {\em is known a priori} not be  $1/f$, e.g. in the Lorenz attractor case where the low frequency spectrum is exponential.

\subsection{``1/f" spectra}

The term $1/f$ spectrum is usually used to denote periodograms where the spectral density $S'(f)$ has an inverse power law form, e.g. the definition used in \cite{Mandelbrot1965b,Mandelbrot1967}
\begin{equation}
S'(f) \sim f^{\theta-2}
\end{equation}
where $\theta$ runs between 0 and 2. 

One needs to distinguish here between bounded and unbounded processes. Brownian, and fractional Brownian, motion are unbounded, nonstationary random walks and one can 
view their  $1/f^{1+2J}$ spectral density as a direct consequence of nonstationarity, as Mandelbrot did (see pp 78-79 of \cite{Mandelbrot1999}). In many physical contexts however, such as the on-off blinking quantum dot process\cite{Niemann2013} or the river Nile minima studied by Hurst\cite{Graves2014b} the signal amplitude is always bounded and does not grow in time, requiring a different explanation that is either stationary or ``conditionally stationary''.

Mandelbrot's best known model for $1/f$ noise remains the stationary, ergodic, fractional Gaussian noise (fGn) that he advocated so energetically in the 1960s. But, evidently himself aware that this had had received a disproportionate amount of attention,  he was at pains  late in his life  (e.g. Selecta Volume N \cite{Mandelbrot1999} p.207, introducing the reprinted \cite{Mandelbrot1965b,Mandelbrot1967}) to stress  that:

\begin{quote}
{\it Self-affinity and an 1/f spectrum can reveal themselves in several {\em quite distinct} fashions ... forms of 1/f behaviour that are predominantly due to the fact that a process does not vary in ``clock time" but in an ``intrinsic time" that is fractal. Those 1/f noises are called ``sporadic" or ``absolutely intermittent", and can also be said to be ``dustborne" and ``acting in fractal time". }
\hfill 
\end{quote}

He clearly distinguishes the LRD stationary Gaussian models like fGn from from his ``conditionally stationary'' fractal time process, noting also that:

\begin{quote}
{\it There is a sharp contrast between a highly anomalous (``non-white") noise that proceeds in ordinary clock time and a noise  whose principal anomaly is that it is restricted to fractal time. }
\hfill 
\end{quote}
 
 In practise the main importance of this is to caution that, used on its own, even a very sophisticated approach to the periodogram like the GPH method \cite{Beran2013} cannot tell the difference between a time series being stationary LRD of the fGn type and ``just" a ``1/f" noise, unless  independent information about stationarity is also available.  
 
One route to reducing the ambiguity in future studies of $1/f$ is to develop non-stationary extensions to the Wiener-Khinchine theorem. An important step \cite{Leibowich2015} has been to distinguish between one which relates the spectrum and the ensemble average correlation function, and a  second relating the spectrum to the time average correlation function. The importance of this distinction can be seen by considering the Fourier inversion of the power spectrum-does the inversion yield the time or the ensemble average?  [E. Barkai, personal communication].

\subsection{LRD}
My readers will, I hope, now be able to see why I believe that the commonly used  spectral definition of LRD has caused misunderstandings. The problem has been that on its own a ``1/f" behaviour is necessary but not sufficient, and stationarity is also essential  for LRD in the sense so widely studied in statistics community (e.g. in \cite{Beran1994} and \cite{Beran2013}). One may in fact argue that the more crucial aspect of LRD is thus the ``loose" one embodied in its name, rather than the formal one embodied in the spectral definition, because {\em a $1/f$ spectrum can  only be synonymous with LRD when there is an infinitely long past}.   The fact that fGn exhibits LRD  {\em by construction} because the stationarity property is assumed, and {\em also} shows   1/f noise,   and the Hurst effect has led to the widespread misconception that the converse is true, and that observing ``1/f" spectra and/or the Hurst effect must {\em imply} LRD.

\section{Conclusions}
Unfortunately \cite{Mandelbrot1967}  received far less contemporary attention   than did Mandelbrot's papers on heavy tails in finance in the early 1960s or the series with van Ness and Wallis in 1968-69 on stationary fractional Gaussian models for LRD, gaining only about 20 citations in its first 20 years. This has been rectified since but I believe the consequences have been lasting.  Perhaps it was because his work on the AFRP was communicated primarily in the (IEEE) journals and conferences of telecommunications and computer science, that it was largely invisible to the contemporary audience that encountered fGn and fBm in SIAM Review and Water Resources Research. In fact, so invisible was it that one his most articulate critics, hydrologist, Vit Kleme\v{s} \cite{Klemes1974} used an AFRP model as a paradigm for the {\em absence} of the type of LRD seen in the stationary fGn model, clearly unaware of Mandelbrot's work. Kleme\v{s}' paper remains very worthwhile reading even today. It showed how at least two other classes of model could exhibit the Hurst effect, the AFRP class and also integrated processes, such as AR(1) with a  $\phi$ parameter close to 1. Fascinatingly, despite the fact that Mandelbrot had colleagues such as the hydrologist Wallis who published in Water Resources Research, and thus may well have seen the paper, if he did he chose not to enlighten  Kleme\v{s} about his earlier work. Sadly Kleme\v{s} and Mandelbrot  seem also not to have subsequently debated nonstationary approaches on an equal footing with fGn,  as  with the advantage of historical distance one can see the importance of both as non-ergodic and ergodic solutions to the $1/f$ paradox.

Although he revisited the 1963-67 fractional renewal papers with  new commentaries in the volume of his Selecta \cite{Mandelbrot1999} that dealt with multifractals and ``$1/f$'' noise, Mandelbrot himself neglected to mention them explicitly in his popular historical account of the genesis of LRD in \cite{Mandelbrot2008}. That he saw them as a representing a different strand of his work to fractional Brownian motion is clear from the way that fBm and fGn and the Gaussian paths to $1/f$ were each allocated a separate Selecta volume \cite{Mandelbrot2002}.  Despite the Selecta, the relatively low visibility has remained to the present day. Mandelbrot's fractional renewal papers are for example not cited or discussed even in encyclopedic books on LRD such as Beran et al's \cite{Beran2013}.

The long term consequence of this in the physics and statistics literatures may have been to emphasise ergodic solutions to the $1/f$ problem at the expense of  non-ergodic ones. This seems to me to be important, because, for example, Per Bak's paradigm of Self-Organised Criticality, in which stationary spectra and correlation functions play an essential role, could not surely have been positioned as the unique solution to the $1/f$ problem \cite{Watkins2016} if it had been widely recognised how different Mandelbrot's two existing routes to 1/f already were.

\paragraph{Acknowledgements.}I would like to thank Rebecca Killick for inviting me to talk at ITISE 2016, and helpful comments on the manuscript from Eli Barkai. I also  gratefully acknowledge many valuable discussions about the history of LRD and weak ergodicity breaking with Nick Moloney, Christian Franzke, Ralf Metzler, Holger Kantz, Igor Sokolov, Rainer Klages, Tim Graves, Bobby Gramacy, Andrey Cherstvy, Aljaz Godec, Sandra Chapman, Thordis Thorarinsdottir, Kristoffer Rypdal,Martin Rypdal, Bogdan Hnat, Daniela Froemberg, and Igor Goychuk among many others. I acknowledge travel support from KLIMAFORSK project number 229754 and the London Mathematical Laboratory, a senior visiting fellowship from the Max Planck Society in Dresden, and Office of Naval Research NICOP grant NICOP - N62909-15-1-N143 at Warwick and Potsdam.

% ---- Bibliography ----
%

\clearpage
%\addtocmark[2]{Author Index} % additional numbered TOC entry
%\renewcommand{\indexname}{Author Index}
%\printindex
\clearpage
%\addtocmark[2]{Subject Index} % additional numbered TOC entry
%\markboth{Subject Index}{Subject Index}
%\renewcommand{\indexname}{Subject Index}
%\input{subjidx.ind}
\end{document}